\documentclass[aps,pra,preprint,amsmath,titlepage,a4paper]{revtex4-2}

\usepackage{braket}
\usepackage{graphicx}

\begin{document}


\title{Experimental requirements for entangled two-photon spectroscopy} 



\author{Stefan Lerch}
\author{André Stefanov}
\email[]{andre.stefanov@iap.unibe.ch}
\affiliation{Institute of Applied Physics, University of Bern, Switzerland}


\date{\today}

\begin{abstract}
Entangled two-photon spectroscopy is expected to provide advantages compared with classical protocols. It is achieved by coherently controlling the spectral properties of energy-entangled photons. We present here an experimental setup that allows the spectral shaping of entangled photons with high resolution. We evaluate its performances by detecting sum frequency generation in a non-linear crystal. The efficiency of the process is compared when performed with classical or entangled light.
\end{abstract}


\maketitle 

\section{Introduction}

\begin{figure}[hbt]
\centering
\includegraphics[width=.2\columnwidth]{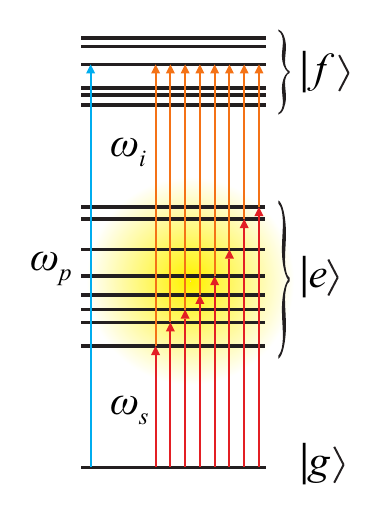}
\caption{Schematic of a two-photon transition with entangled photons. Because signal and idler are broadband, they access the entire manifold $\ket{e}$. At the same time their frequencies sum up to the frequency of the pump that can originate from a monochromatic laser ($\omega_p=\omega_s+\omega_i$), and thus the manifold $\ket{f}$ is well resolved.}
\label{fig:TPAlevelscheme}
\end{figure}

Optical quantum metrology aims at realizing novel measurement schemes in which classical light is replaced by quantum states of light. It is expected to offer improved sensitivity by overcoming the shot-noise limit, or even allowing for completely new procedures. For the spectroscopic investigation of two photon processes, it was theoretically shown that time-energy entangled states of light provide  advantages compared to classical spectroscopy methods (see \cite{Dorfman2016a} for a review). The most efficient source of entangled light for the time being is spontaneous parametric downconversion (SPDC) in non-linear crystals \cite{Couteau2018}. In particular, energy-entangled photons are generated pairwise in the SPDC process. Therefore, such source of light is well suited to probe two-photon processes with entangled states, such as two-photon absorption (TPA). TPA with energy-time entangled photons shows two obvious features. First, the pairwise appearance at the sample leads to a linear dependence of the absorption signal on the incoming photon flux \cite{Gea-Banacloche1989,Javanainen1990,Georgiades1995,Fei1997,Lee2006a,Upton2013}. As a consequence, the power required for TPA can be reduced considerably for continuous wave excitation. Second, energy-time entanglement leads simultaneously to high frequency and time resolution. This has been experimentally shown for an off-resonant two-photon transition in rubidium by Dayan et al. \cite{Dayan2004}. The feature can be applied to selectively probe two-photon transitions to double-excitation states \cite{Schlawin2012b,Schlawin2013a,Oka2011a,Oka2011b} (see Fig. \ref{fig:TPAlevelscheme}). Whereas the energies of the two photons of the pair sum up to the energy of the pump photon, which can originate from a monochromatic laser, the individual spectrum of each photon can be very broad. Other effects associated with entangled TPA are entanglement-induced transparency \cite{Fei1997} and virtual-state spectroscopy \cite{Saleh1998,Kojima2004}. The last three references introduced nonlinear spectroscopy with entangled photons with new control parameters, such as the delay between signal and idler photon, and the so-called entanglement time.

In this article, we investigate the experimental requirements to implement two-photon spectroscopic schemes with entangled light. At first we present an experimental setup to manipulate the energy parameters of a two-photon state with high resolution. A grating compressor combined with a spatial light modulator allow to control the transmission and phase of each frequency component of the states. The detection is realized by sum frequency generation (SFG). We evaluate the energy resolution and tuneablity of the setup in the perspective of nonlinear spectroscopy with entangled photons. We further experimentally estimate the enhancement of the SFG rate using entangled photons with the rate obtained with uncorrelated photons. The formal equivalence between SFG and TPA allows to analyze the feasibility of observing TPA with the grating compressor setup. 

\section{Experimental setup}

\subsection{Concept}
\label{sec:CC_concept}
One of the important tools of ultrafast optics are pulse shapers, that consist in a combination of a pulse compressor with a spatial light modulator (SLM). They allow to generate nearly arbitrarily shaped optical waveforms \cite{Weiner2000a}. Pulse shaping had particularly large influence on coherent control of quantum systems \cite{Rice200Book}. Inspired by femtosecond pulse shaping, the Silberberg group was first to applied the concept of ultrafast optics to the energy-time shaping of quantum states of light \cite{Dayan2005,Pe2005}.

The basic idea of a shaping setup is to control the temporal shape of a laser pulse. An incident laser pulse is described by its slowly varying envelope $\mathcal{E}_{in}(\Omega)$ in the frequency domain with a frequency $\Omega=\omega-\omega_c$ relative to a center frequency $\omega_c$. By means of a dispersive element and an SLM, it is possible to control the phase and/or amplitude of each frequency component. The action of the pulse shaper is described by a transfer function $M(\Omega)$ such that the outgoing pulse reads
\begin{equation}
\mathcal{E}_\mathit{out}(\Omega)=\mathcal{E}_\mathit{in}(\Omega)M(\Omega).
\end{equation}
In quantum optics the field operator is spectrally modified \cite{Shih2020book}.
\begin{equation}
\hat{E}(\Omega)^+\propto\hat{a}(\Omega) \quad\rightarrow\quad \hat{E}(\Omega)^+\propto\hat{a}(\Omega)M(\Omega)
\end{equation}
Formally equivalent, the shaping process can be seen as a manipulation of the photon's energy modes
\begin{equation}
\ket{\Psi}=\int\mathrm{d}\Omega\;S(\Omega)\ket{\Omega}\quad \rightarrow\quad \ket{\Psi}=\int\mathrm{d}\Omega\;S(\Omega)M(\Omega)\ket{\Omega}
\end{equation}

\begin{figure}[htbp]
\centering
\includegraphics[width=0.8\linewidth]{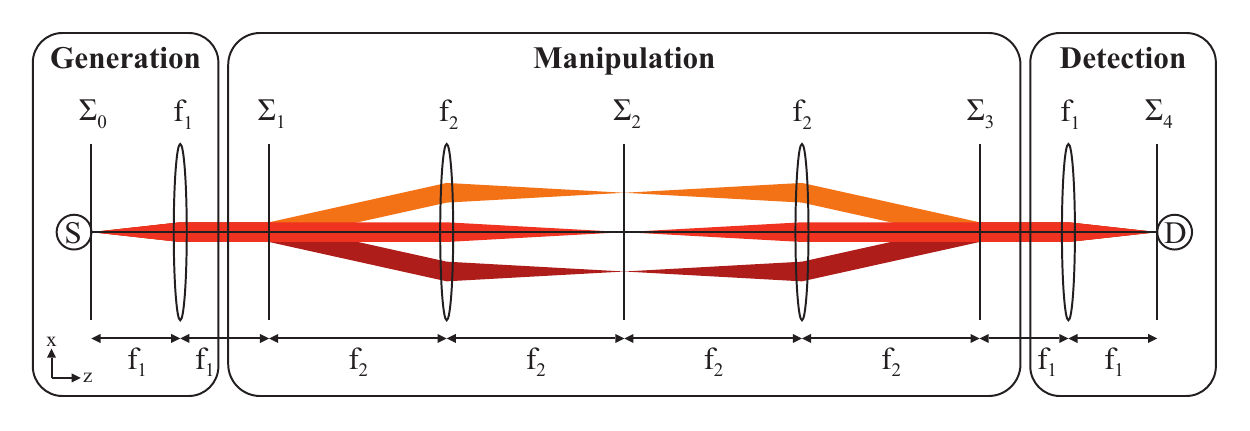}
\caption{Principle of the setup. The light generated by the source S at the plane $\Sigma_0$ is imaged by two lenses of focal length $f_1$ and $f_2$ to $\Sigma_2$. A dispersive element at $\Sigma_1$ separates the spectrum spatially. At $\Sigma_2$, an SLM shapes the frequency modes of the light. The setup is then folded such that $\Sigma_3$ is an image of $\Sigma_1$, and the plane of detection (D, $\Sigma_4$) is an image of $\Sigma_0$.}
\label{fig:setup_GC}
\end{figure}

We can conceptually subdivide the experimental setup into three parts, as illustrated on Figure \ref{fig:setup_GC}: Generation of energy-time entangled photons, manipulation of the frequency modes and detection by means of some nonlinear process.

\subsection{Energy entangled photons}
SPDC appears in a crystal with non-zero second-order susceptibility $\chi^{(2)}$. Occasionally, the pump photon ($p$) gets annihilated and two photons, signal ($s$) and idler ($i$), are created. We denote $\omega_j$ as the angular frequency, and $\mathbf{q}_j$ as the transverse part of the wave vector $\mathbf{k}_j$, $j\in\{p,s,i\}$. Using time-dependent perturbation theory in the interaction picture up to first order, an entangled two-photon state generated by SPDC reads
\begin{equation}
\ket{\Psi}=\ket{0}+\int\mathrm{d}^2q_i\int\mathrm{d}\omega_i\int\mathrm{d}^2q_s\int\mathrm{d}\omega_s\;\Lambda(\mathbf{q}_i,\omega_i,\mathbf{q}_s,\omega_s)\ket{\mathbf{q}_i,\omega_i}_i\ket{\mathbf{q}_s,\omega_s}_s.
\label{eq:Psi}
\end{equation}
The joint spectral amplitude (JSA)
\begin{equation}
\Lambda(\mathbf{q}_i,\omega_i,\mathbf{q}_s,\omega_s)=\mathcal{E}^+_p(\mathbf{q}_i+\mathbf{q}_s,\omega_i+\omega_s-\omega_p)f_{DC}(\mathbf{q}_i,\omega_i,\mathbf{q}_s,\omega_s)
\label{eq:SPDC_Lambda}
\end{equation}
is the product of a pump envelope function $\mathcal{E}^+_p(\mathbf{q}_i+\mathbf{q}_s,\omega_i+\omega_s-\omega_p)$ and a phase-matching function $f_{DC}(\mathbf{q}_i,\omega_i,\mathbf{q}_s,\omega_s)$. In the transverse direction the crystal dimensions are much larger than the waist of the pump beam. This leads to perfectly phase-matched transverse momentum \cite{Shih2003}.
\begin{equation}
\mathbf{q}_p=\mathbf{q}_i+\mathbf{q}_s.
\label{eq:qpqiqs}
\end{equation}
In this article, a monochromatic pump
\begin{equation}
\mathcal{E}^+_p(\mathbf{q}_i+\mathbf{q}_s,\omega_i+\omega_s-\omega_p)=\mathcal{E}^+_p(\mathbf{q}_i+\mathbf{q}_s)\delta(\omega_i+\omega_s-\omega_p)
\label{eq:SPDC_Epump}
\end{equation}
is used in order to generate highly energy-time entangled photons. This frequency correlation makes $\Lambda(\mathbf{q}_i, \omega_i, \mathbf{q}_s, \omega_s)$ non-separable with respect to the variable pair $(\omega_i,\omega_s)$, and therefore, the state $\ket{\Psi}$ is called energy-time entangled.

\begin{figure}[htbp]
\centering
\includegraphics[width=0.6\linewidth]{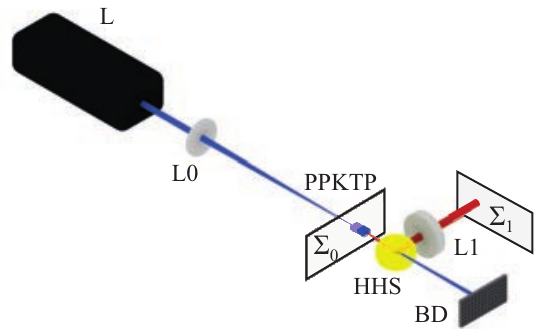}
\label{fig:SPDC_generation}
\caption{Sketch of the SPDC source. A pump laser (L) is focused by a lens (L0) into a PPKTP crystal at $\Sigma_0$. By means of SPDC, energy-time entangled photons are created and separated from the pump by a high harmonic separator (HHS). The residual pump is blocked by a beam dump (BD). The down-converted light is collimated by lens L1 and coupled into the manipulation part. The plane $\Sigma_1$ is a $2f_1$-image plane of $\Sigma_0$.}
\end{figure} 

The generation setup is sketched in Fig. \ref{fig:SPDC_generation}. As a pump laser, we use a frequency doubled Ti:Sa laser from HighFinesse. It consists of an 8 W Finesse (Laser Quantum), which pumps a Ti:Sa crystal in a Z-fold geometry. The laser can be stabilized to two or three modes that are separated by the free spectral range of the cavity $\Delta\nu_\mathit{FSR}=2$ GHz. The free running laser has a linewidth which is smaller than 0.35 MHz (measured over 50 ms). We chose the center wavelength to be $\lambda_\mathit{Ti:Sa}=800$~nm, and a maximum output power between 1.8 to 2.1 W. The Ti:Sa beam is sent into a frequency doubling stage from HighFinesse. In a ring cavity which is resonant for $\lambda_\mathit{Ti:Sa}$, a potassium dihydrogen phosphate KH$_2$PO$_4$ (KDP) crystal generates s-polarized light (electric field perpendicular to the optical table) centered around $\lambda_p=\lambda_{\mathit{Ti:Sa}}/2=400$ nm by means of type-I SFG. While the nominal output power amounts $P=0.8$ W, the usual output power is $P=0.5$ W.

The rotational asymmetry of the beam is corrected by a cylindrical lens and an N-KZFS8 anamorphic prism pair. The remaining light to pump the SPDC process has typically $0.28$ W power. 

\begin{figure}[t]
\centering
\includegraphics[width=0.5\columnwidth]{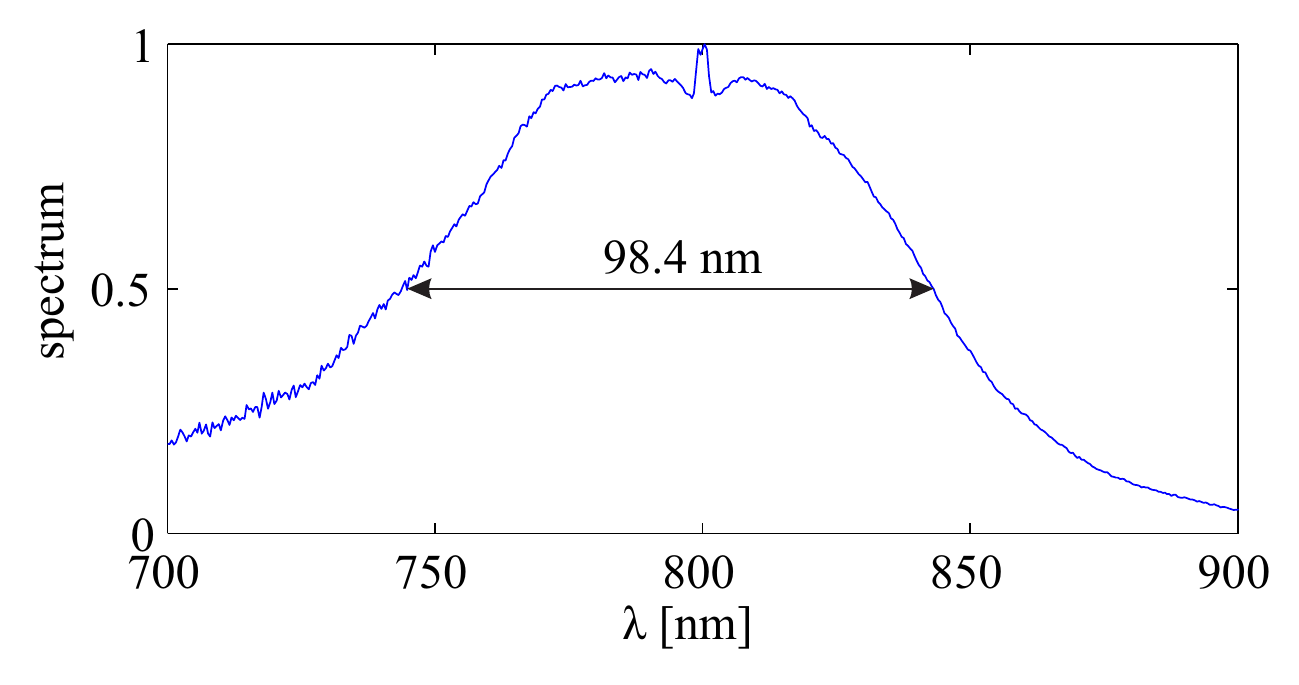}
\caption{Down-converted spectrum normalized to one. The light is collimated with a lens after the SPDC crystal ($50$ mm focal length) and coupled with a collimator into a fiber coupled spectrometer with $2$ nm resolution. The peak at $800$ nm stems from some residual pump light ($400$ nm) that is detected because of second-order diffraction in the spectrometer.}
\label{fig:CC_spec}
\end{figure}

The pump is focused by lens L0 of focal length $150$ mm into a PPKTP crystal (plane $\Sigma_0$). The focal intensity amounts approximately $22$ kW/cm${}^2$. The crystal is $12$ mm long, has $3.5$ \textmu m poling period, and is mounted in a copper block. An indium foil surrounds the crystal and guarantees heat transfer between crystal and copper. The copper block is temperature stabilized by a water heating system up to $0.1^\circ$C. The down-converted photons are degenerated at $\lambda_c=800$ nm. They are broadband with a FWHM of about $98$ nm (see Fig. \ref{fig:CC_spec}). The down-converted power of $P = 200$ nW translates to a spectral mode density of $n = 0.017$. We chose the phase-matching for mostly collinear emission at $\lambda_c$ in type-0 configuration, i.e. all involved photons have the same polarization (s-polarization). A high harmonic separator (HHS) separates the down-converted light from the residual pump beam. The pump is blocked by a beam dump. A lens L1 of focal length $f_1=50$ mm collimates the down-converted light. After the distance $f_1$, at plane $\Sigma_1$, signal and idler enter the manipulation stage.

\subsubsection{Wavelength Tunability}
\label{sec:CC_tunability}

\begin{figure}[b]
\centering
\includegraphics[width=0.6\linewidth]{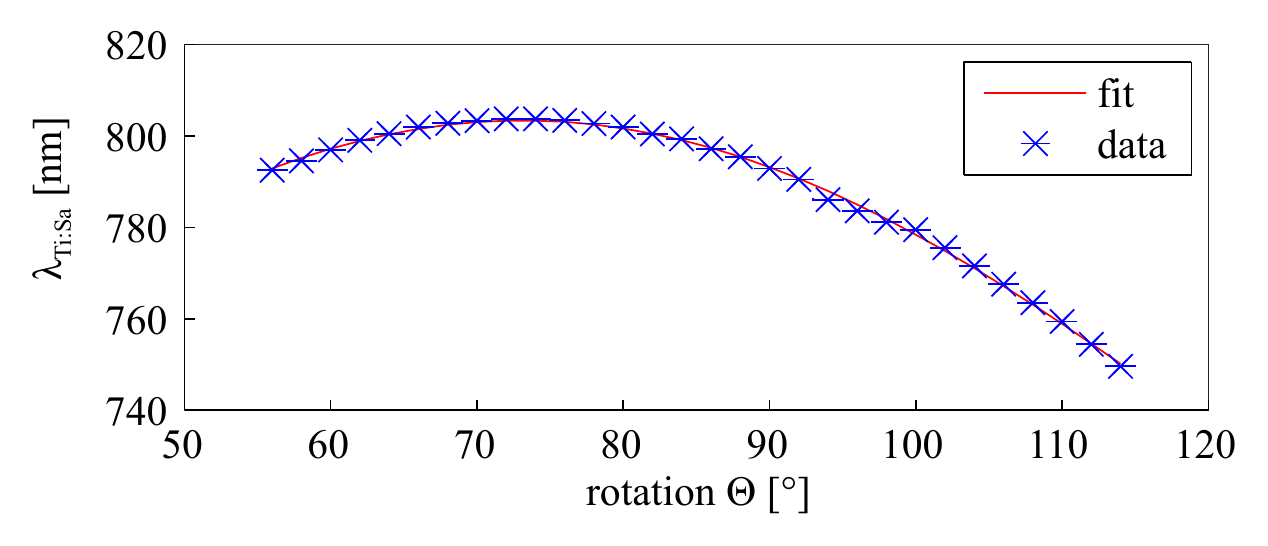}
\caption{By rotating the interference filter in the Ti:Sa cavity, the wavelength can be tuned over a large wavelength range. The data (blue crosses) have some uncertainty of the rotation angle, indicated horizontally. The wavelength is measured by a wavemeter (WA-20, Burleigh) with a helium neon laser as reference. The fit (red curve) is given by the properties of the filter \protect\cite{Baillard2006}.}
\label{fig:GC_pump_wavelength}
\end{figure} 

Towards the use of entangled photons for spectroscopy, one of the parameters that have to be tuned is their central frequency. The Ti:Sa laser covers from $700$ nm to $1100$ nm a wide spectrum of emission wavelengths. By means of an interference filter in the cavity the emission wavelength can be continuously selected in the range from $770$~nm up to $803$~nm, as Fig. \ref{fig:GC_pump_wavelength} illustrates. Furthermore, a frequency stabilization of the cavity to an external reference, e.g. an atomic transition, could be implemented over a range of 10 GHz, in order to reduce the linewidth to less than $60$ kHz rms. The stabilization is not implemented here because the phase-matching of pump-frequencies in the nonlinear crystal is much broader.

The Ti:Sa beam gets frequency doubled in a KDP crystal. The phase-matching curve of this process is relatively narrow, such that the crystal tilt and the cavity have to be realigned for wavelength tuning over more than one nanometer. However, with realignment, the crystal cut allows to phase-match the SHG process in the range from $380$ nm to $410$ nm.

The phase-matching condition becomes even more critical for the SPDC process in the PPKTP. A change in the pump wavelength behaves similarly to a change in the poling period $G$ or in the crystal temperature $T$. Decreasing the pump wavelength has the same effect as increasing the temperature with a linear dependence $\partial\lambda_{p}/\partial T=(-0.019\pm0.001)$ nm/${}^\circ$C. This means that one crystal with a fixed poling period can only be used for a $\lambda_p$-change of about one nm, given that the crystal heating system is able to tune the temperature in a range of $52^\circ$C. More tunability in the pump wavelength requires other poling periods of the crystal.

In summary, the setup allows for tuning without realignment within the small range of approximately one nm. This would be enough to scan along atomic transitions. A broader tuning range can be accessed too, but it involves realignment of optics and the change of the SPDC crystal because the phase-matching condition limits the range considerably.

\subsection{Energy mode shapper}

\label{sec:CC_Grating_compressor}
Previous experiments on shaping of entangled photons have been realized with pulse shapers where the dispersive elements are prisms leading to limited energy resolution \cite{Pe2005,Zah2008,Bessire2014}. In order to overcome the limitation of a prism compressor setup with regards to spectroscopic experiments, we exploit the fact that gratings have a higher dispersion than prisms.

\begin{figure}[t]
\centering
\includegraphics[width=0.52\linewidth]{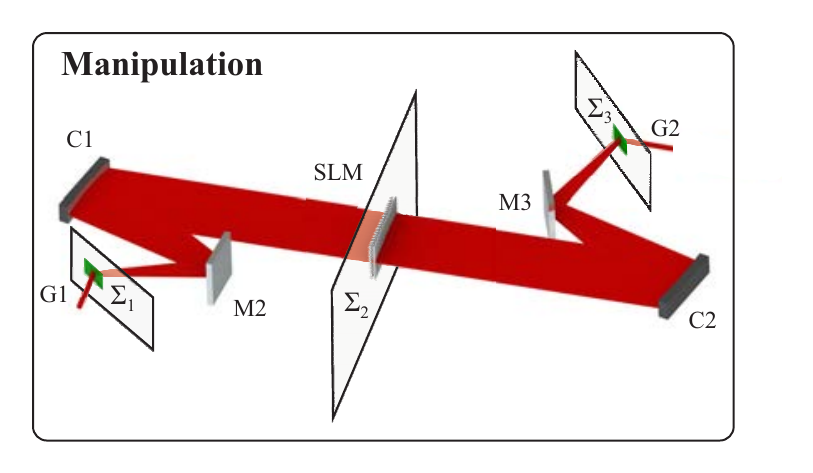}
\caption{Sketch of the manipulation part. It is a grating compressor in a Z-geometry, consisting of two gratings (G1 and G2), two plane mirrors (M2 and M3), and two cylindrical mirror (C1 and C2). At the symmetry plane $\Sigma_2$, the frequency modes are spatially separated, and an SLM allows to shape them in amplitude and phase.}
\label{fig:GC_manipulation}
\end{figure}

The setup of the two-photon shaper is sketched in Fig. \ref{fig:GC_manipulation}. It consists of two gratings (G1 and G2), two plane mirrors (M2 and M3), two cylindrical mirrors (C1 and C2), and the SLM. All elements are aligned in a Z-geometry within a plane ($x$ direction).

The light from the generation stage enters the grating compressor at plane $\Sigma_1$, where a grating disperses the spectral components of the light. We use transmission gratings from LightSmyth with $1503.76$ lines/mm (grating constant $G_1=1/1503.76$ mm), diffraction order $m_1=1$, incidence angle $\alpha_1=41^\circ$, diffraction angle $\beta_{1}=33.2^\circ$ at center frequency $\omega_c=2\pi c/\lambda_c$, and linearized dispersion coefficient $\gamma=2\pi m_1/[\omega_cG_1\cos(\beta_1)]=4.8$~fs/\textmu m. We chose transmission gratings rather than reflection gratings because of the lower losses. The transmission is higher than 90\% for the first diffraction order in the wavelength range $\lambda\in$ [$750$ nm, $850$ nm]. A high transmission of the overall setup is crucial because the coincidences scale quadratically with the single photon losses. The grating changes the orientation of the optical axis, and therefore, scales the beam diameter in $x$ direction by a factor of $b=\cos(\alpha_1)/\cos(\beta_1)$.

The cylindrical mirror C1 of curvature $R=600$ mm is placed in the distance $f_2=R\cos(\theta_c)/2=298.4$~mm away from $\Sigma_1$. The angle $\theta_c=6^\circ$ is the incidence angle of frequency mode $\omega_c$ at C1. Mirror M2 reduces $\theta_c$ in order to suppress astigmatism. The cylindrical mirror C1 collimates all frequency modes with respect to each other, and focuses the transverse beam profile of each frequency mode to $\Sigma_2$. Including the beam diameter scaling $b$, one finds a magnification $M=bf_2/f_1=5.4$ from plane $\Sigma_0$ to $\Sigma_2$.

At the symmetry plane $\Sigma_2$, the spatially dispersed spectrum passes two identical nematic liquid crystal arrays of a programmable SLM (Jenoptik SLM-S640d USB). Each array consists of $640$ single pixels with $97$ \textmu m in width and a gap of $3$ \textmu m between two adjacent pixels. The SLM in combination with the polarization-dependent detection method allows to shape the photons phase and amplitude independently. The grating and the mirror curvature is chosen such that the $640$ pixel of the SLM are illuminated by a spectrum of approximately $740$ nm to $860$ nm.

From $\Sigma_2$ to $\Sigma_3$, the spatially dispersed spectrum is recombined in the opposite way as it has been dispersed. The cylindrical mirror C1 and C2 are identical. At distance $f_2$ from $\Sigma_2$, C2 focuses the frequencies to the grating G2 and collimates the spatial beam profile of each frequency mode again. Mirror M3 allows C2 to be tilted by angle $-\theta_c$ with respect to the incidence beam with frequency mode $\omega_c$. Mirror C1 and C2 form essentially a $4f_2$ image from $\Sigma_1$ to $\Sigma_3$, but only in $x$ direction. In $y$ direction the beam propagates freely from $\Sigma_1$ to $\Sigma_3$. That is why it is important to enter the manipulation stage with a collimated beam. Grating G2 is identical to G1, except for the alignment parameters, i.e. the incidence angle $\alpha_2=\beta_1$ and diffraction angle $\beta_2=\alpha_1$ are interchanged. The alignment of the shapper is performed with a classical broadband source (see Appendix \ref{app:alignement}) and the overall transmission is ($62\pm 1$)~\%.

\subsection{Detection}
At the plane $\Sigma_3$, the shaped light enters the detection stage. After the distance $f_1$ (focal length of lens L2), the beam is focused into the center of a nonlinear crystal NLC2 at $\Sigma_4$. The crystal is a $12$ mm long PPKTP, identical to the down-conversion crystal up to fabrication variations. It is identically mounted and the temperature is stabilized with a similar water heating system. In the crystal, the light occasionally undergoes up-conversion. At distance $f_3$, the up-converted beam is collimated by the lens L3 (focal length $f_3=60$ mm). Two filters (shortpass and bandpass, F) separate the up-converted photons from the residual light. Finally, the light is detected by one of three different possibilities. If the light is generated with the down-conversion source, a lens L4 of focal length $f_4=19$ mm focuses the light onto a single photon counting module (id100-50 from ID Quantique). Alternatively for classical SFG, the light is detected with a photodiode (PD-300 sensor from Ophir), or coupled into a fiber and analyzed with a spectrometer (AvaSpec-256 from Avantes, spectral range $350$ nm to $450$ nm).

\label{sec:GC_detection}
\begin{figure}[t]
\centering
\includegraphics[width=0.4\linewidth]{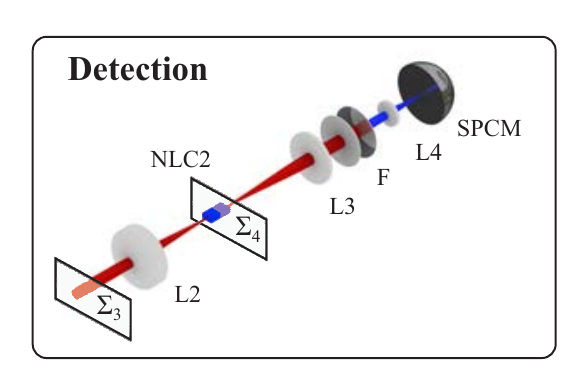}
\caption{Sketch of the detection part. The light is detected by up-conversion in the nonlinear crystal NLC2. The up-converted photons are imaged by lens L3 and L4 onto a single photon counting module (SPCM). A filter set, consisting of a lowpass and bandpass filter (F), separates the up-converted photons from the residual light.}
\label{fig:GC_detection}
\end{figure}

\subsection{Energy Resolution}
\label{sec:CC_resolution}
A good spectral resolution at $\Sigma_2$ is essential for reliable energy mode shaping. There are essentially three quantities that characterize the resolution: the change of center frequency per space unit, the point spread function, and the space that is illuminated by one wavelength.

\begin{figure}[t]
\centering
\includegraphics[width=1\linewidth]{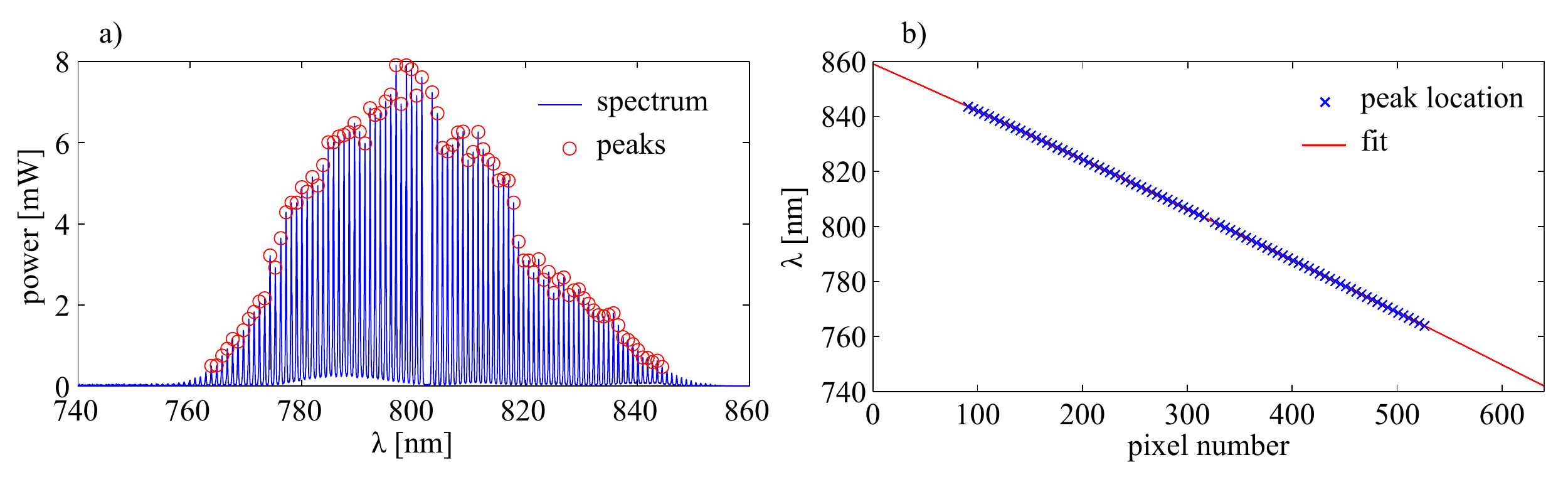}
\caption{a) Spectrum of the classical source for the wavelength calibration of the SLM. Every fifth pixel rotates the polarization of the incoming light such that the passing spectral components are detected. The gap at pixel $321$ allows to uniquely identify each peak (circles) to a pixel. b) Employing the fitting model of Eq. \eqref{eq:CC_wavlength_to_pixel} on the peak location data results in the mapping of a central wavelength for each SLM pixel.}
\label{fig:GC_spectral_calibration}
\end{figure}

Starting from $\Sigma_0$ where ideally all frequency modes spatially overlap, the imaging system and the dispersive element determine the change of center frequency per space unit. We chose the space unit to be the SLM's pixel width of $\Delta x=100$ \textmu m and measure $(0.541\pm 0.001)$ mrad/(fs pixel) for the pixels around $\lambda_c$. The space to frequency mapping is not completely linear. This can be seen in the wavelength to pixel calibration, exemplarily plotted for the classical source in Fig. \ref{fig:GC_spectral_calibration}. The spectral calibration is equal for both sources, since the optical axes of the center frequency are the same. In the experiment, the deviation is typically within one pixel. A polarizing beam splitter is oriented such that no light is coupled into a multimode fiber with $100$ \textmu m core diameter at the position of the up-conversion crystal. With the SLM, we rotate the polarization at every fifth pixel, i.e. the light passing through the affected pixels can be selected. Subsequently, it is analyzed by an OSA with 0.1~nm resolution (Fig. \ref{fig:GC_spectral_calibration} a). For each ``opened'' pixel, we measure a corresponding peak in the spectrum. Pixel 321 is not opened, thus we can uniquely identify a pixel number to each peak (Fig. \ref{fig:GC_spectral_calibration} b). The directionality of the calibration is determined previously by rotating the polarization only at pixels 1 to 320. The fitting model is derived from the grating equation \cite{Martinez1986}
\begin{equation}
\beta(\lambda)=\textrm{asin}\left(\frac{m\lambda}{G}-\textrm{sin}(\alpha)\right),
\label{eq:CC_grating_equation}
\end{equation}
with the diffraction order $m=1$, grating constant $G$ as fitting parameter around 0.67 1/\textmu m, input angle $\alpha = 41^\circ$, and the diffraction angle $\beta(\lambda)$ that is linked with the pixel number $p$ by 
\begin{equation}
\tan\left(\beta(\lambda)-\beta(\lambda_c)\right)=\frac{p-p_0}{f_2}\Delta x.
\label{eq:CC_wavlength_to_pixel}
\end{equation}
Solving Eq. \eqref{eq:CC_wavlength_to_pixel} for $\lambda$, yields
\begin{equation}
\lambda(p)=G\left\{\sin\left[\textrm{atan}\left(\frac{p-p_0}{f_2}\Delta x\right)
+\beta(\lambda_c)\right]
+\sin\left(\alpha\right)\right\},
\end{equation}
with fitting parameters $G$, the center pixel $p_0$, and the focal length $f_2$.

\begin{figure}[t]
\centering
\includegraphics[width=1\linewidth]{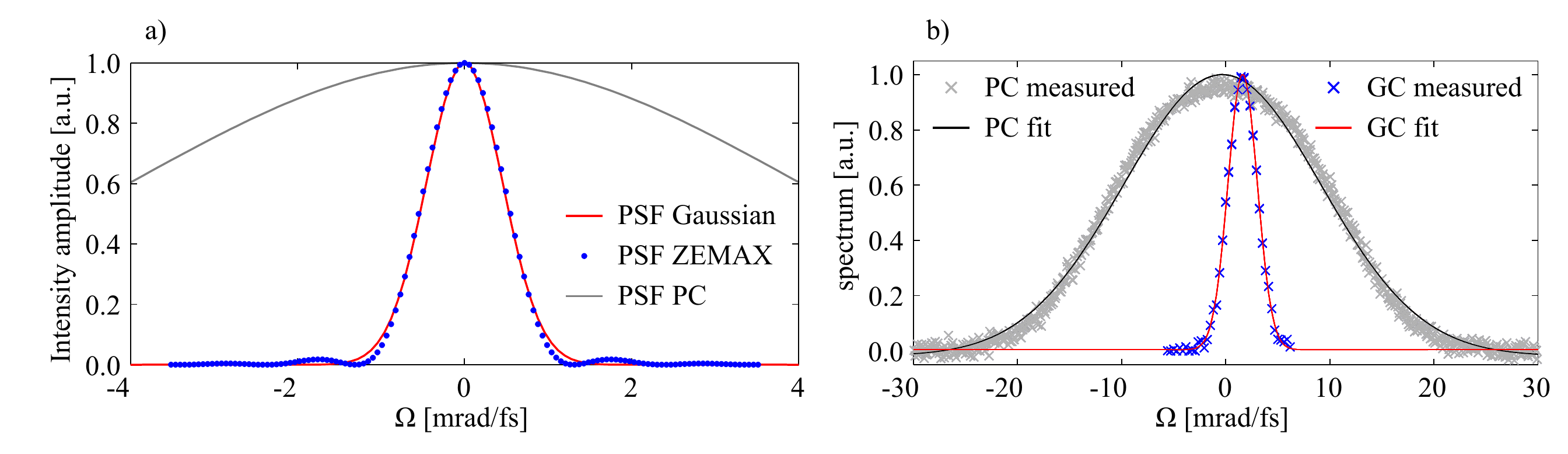}
\caption{a) The PSF according to Zemax (blue dots) is approximated by a Gaussian function (red curve) of (FWHM) $\Delta\Omega_\mathit{PSF}= 0.756\pm 0.006$ mrad/fs. The gray curve indicates the PSF of the prism compressor (PC) setup. b) Normalized spectra that pass the five central pixels of the grating compressor (GC, blue crosses) and prism compressor (PC, gray crosses) setup. Both spectra are fitted by Gaussian functions of FWHM $(22.4\pm 0.2)$~mrad/fs (PC, black curve) and $(3.30 \pm 0.06)$~mrad/fs (GC, red curve).}
\label{fig:GC_PSF}
\end{figure} 

The second resolution parameter, the point spread function (PSF), is calculated using the Huygens method in ZEMAX. The result is plotted in Fig. \ref{fig:GC_PSF} a). The result is well approximated by a Gaussian function with FWHM of $(0.756\pm0.006)$~mrad/fs. It is about nine times smaller than the one of the prism compressor \cite{Bessire2014}.

The PSF describes the spreading of a point-like source. In reality, the source has a finite beam waist $w_0$, which leads to the third resolution parameter, the space that is illuminated by one wavelength. The imaging system magnifies the beam waist $w_0$ of one frequency mode at $\Sigma_0$ to $\Sigma_2$ by a factor of $M=5.4$. Therefore, the down-conversion beam waist of approximately $20$ \textmu m translates to an illuminated space of $216$ \textmu m (two pixels) at the SLM. The PSF blurs this to about five pixels. For a direct comparison between prism and grating compressor, we analyze the inverse parameter, i.e. we measure the spectrum transmitted through five pixels. For both setups, the space of five pixels is just at the limit where one frequency mode gets fully transmitted. Therefore, no flat-top spectrum is observed and the spectra can be fitted with a Gaussian function. For both setups the pixels near the center wavelength ($\lambda_c=800$ nm and $\lambda_c=1064$ nm) are set to transmission one, the others to zero. The spectra [Fig. \ref{fig:GC_PSF} b)] are measured by the OSA with a resolution of $1.0$~nm for the prism compressor, $0.1$~nm for the grating compressor, respectively. The spectra are filtered by a polarizing beam splitter and are coupled into a single mode fiber at the positions of the up-conversion crystal. For comparison, the spectra are normalized to the maximum of the corresponding fit. The FWHM are $(22.4\pm 0.2)$ mrad/fs for the prism compressor and $(3.30 \pm 0.06)$ mrad/fs for the grating compressor setup. The ratio of those widths is $6.8\pm 0.1$, and illustrates again the better frequency resolution of the grating compressor setup.

\section{Requirements for Nonlinear Spectroscopy with Entangled Photons}
\label{sec:TPA_suitable_setup}
The main requirements for a setup to be suited for nonlinear spectroscopy with entangled photons are the ability to obtain a TPA signal high enough and the capability to coherently control the process by shaping the two-photon energy wavefunction, in a way to exploit the simultaneous frequency and time resolution of the source. In this section, we describe those requirements and show to what extend the setup complies with them.

\subsection{Entanglement Time and Entanglement Area}
In general, the TPA rate for continuous sources of light
\begin{equation}
R=\sigma_e \phi+\sigma_c\phi^2
\end{equation}
involves a linear and a quadratic dependence on the incoming flux density $\phi$, for correlated photon pairs $\phi/2$ is the number of pairs. The quadratic term describes the absorption of two photons from different pairs that randomly arrive at the two-photon absorber, i.e. the quadratic term is also present for uncorrelated photons. The two-photon quadratic cross section $\sigma_c$  quantifies the strength for TPA with classical light (randomly arriving photons). Entangled TPA profits from the linear scaling at low flux. Out of any resonances the entangled two-photon cross section is approximately given by \cite{Fei1997}
\begin{equation}
\sigma_e=\frac{\sigma_c}{2A_eT_e}.
\label{eq:TPA_sigmae}
\end{equation}

The entanglement time $T_e$ and entanglement area $A_e$ are parameters that describe the entangled two-photon state at the absorber. In Eq. \eqref{eq:TPA_sigmae}, the entanglement time is regarded as the width of the temporal second-order correlation function, i.e. it describes the time within signal and idler arrive at the two-photon absorber. This should not be confused with the coherence time, given by the width of the second-order correlation function when all phase contributions are zero, i.e. when transform-limited. Similarly, the entanglement area describes the correlation area between the spatial localization of signal and idler photons at the two-photon absorber. Small $T_e$ and $A_e$ are desirable to get a high TPA rate. It basically states that temporally and spatially confined photons enhance the TPA rate. The same is true for classical TPA. Confining the photons in space (by focusing) and time (with pulsed sources) leads to a higher TPA rates.

In the following we derive the relation between the classical and entangled SFG coefficients from an analogy between photon pairs and classical pulses. The classical two-photon response to a flux $\phi_{IR}(\vec x,t)$, assuming an instantaneous response of the medium (as in a thin SFG crystal) is given by
\begin{equation}
\phi_{SFG}(\vec x,t)=\beta_c \phi_{IR}(\vec x,t)^2,
\end{equation}
where the fluxes are expressed in photon/s/m$^2$ and $\beta_c$ in m$^2$s. The total incoming IR and generated SFG intensities are given by 
\begin{equation}
I_{IR,SFG}=\int\phi_{IR,SFG}(\vec x,t)d\vec x.
\end{equation}
For a spatially uniform of area $A$ and continuous IR beam we have
\begin{equation}
I_{SFG}=\frac{\beta_c I_{IR}^2}{A}.
\end{equation}
More generally, for pulses with a Gaussian temporal shape of width $\tau$ and a Gaussian circular spatial distribution of width $\sigma$, we have
\begin{equation}
    \phi_{IR}(\vec x,t)=\frac{N_{IR}}{2\pi\sigma^2\sqrt{2\pi}\tau}\exp{\left(-\frac{|\vec x|^2}{2\sigma^2}\right)}\exp{\left(-\frac{t^2}{2\tau^2}\right)},
\end{equation}
were $N_{IR}$ is the total number of photon in one pulse. The average intensity in photon/s is $I_{IR}=N_{IR}\nu$ with $\nu$ the repetition rate of the laser. The total SFG signal is expressed by
\begin{equation}
    I_{SFG}=\nu\beta_c\int \phi_{IR}(\vec x,t)^2d\vec x dt,
\end{equation}
such that after integrating  
\begin{align}\label{eq:ISFGPulsed}
    I_{SFG}=&\frac{\beta_c\nu}{8\sqrt{\pi}\pi^2\sigma^2\tau}N_{IR}^2\\
    =&\frac{\beta_c}{8\sqrt{\pi}\pi^2\sigma^2\tau\nu}I_{IR}^2.
\end{align}
In the case of a continuous source of photon pairs, the relation is assumed to be linear
\begin{equation}
\phi_{SFG}(\vec x)=\beta_q \phi_{IR}(\vec x).
\end{equation}
The total upconverted intensity is thus $I_{SFG}=\beta_q I_{IR}$.

A relation between $\beta_c$ and $\beta_q$ is obtained by conceptually identifying photon pairs with entanglement time $\tau_e$ and entanglement size $\sigma_e$ to a train of pulses of duration $\tau=\tau_e$, size $\sigma=\sigma_e$ and with intensity $N_{IR}=2$ photon/pulse, at a repetition rate of $\nu_e$, such that the average power is $I_{IR}=2\nu_e$. From Eq. \eqref{eq:ISFGPulsed} we obtain
\begin{equation}
\label{eq:betaq_betac}
    \beta_q=\frac{\beta_c}{4\sqrt{\pi}\pi^2\sigma_e^2\tau_e}
\end{equation}

\begin{figure}[!b]
	\centering
	\includegraphics[width=0.9\columnwidth]{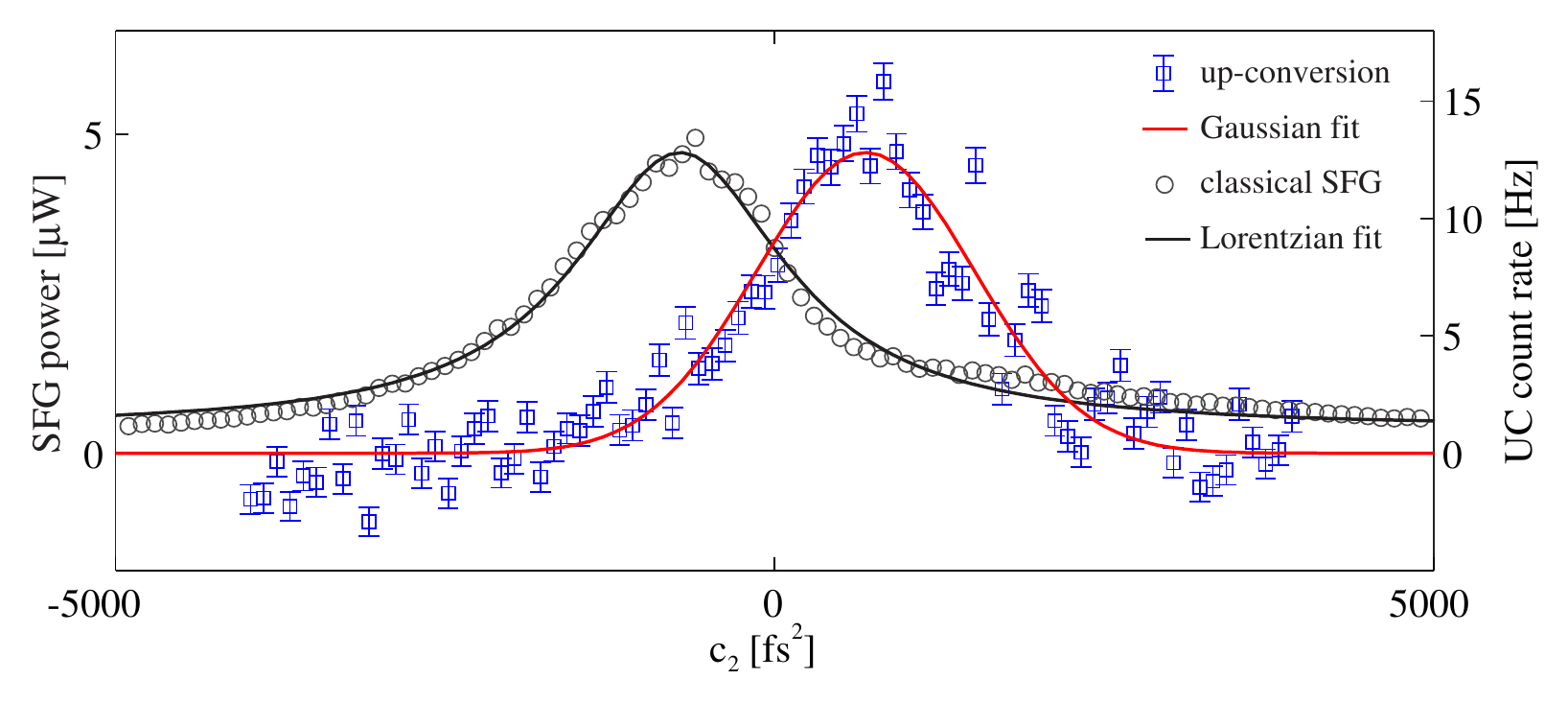}
	\caption{Dispersion scan for the classical source and the quantum source with the grating compressor setup. The scan is done by means of the SLM. The small offset to zero dispersion can easily be compensated by the position of the second grating in the setup. For the quantum source, the measured up-conversion rate (squares) has been fitted by a Gaussian function (red curve) with a maximum of $R_{UC}=(12.8\pm 0.9)$ Hz. The spectral width of the classical source (circles) has been reduced by means of the SLM to match the width of both curves. The fit (black curve) of the measurement follows a Lorentzian function.}
	\label{fig:TPA_c2_scan}
\end{figure}

In order to experimentally test the validity of Eq. \eqref{eq:betaq_betac}, we experimentally compared the SFG signal from the same crystal for classical pulsed light and for entangled photons. SFG in a long crystal is formally equivalent to TPA to a final state with narrow linewidth and no-resonant intermediate states \cite{Dayan2007}. In order to compare classical and quantum TPA, we manipulate the spectral width of the classical source such that the dispersion scan matches the one for the quantum source (see Fig. \ref{fig:TPA_c2_scan}). The scan is done using the SLM with the transfer function according to Eq.\eqref{eq:M_c2}. The entanglement time leading to such dispersion scans is $\tau_e=(24.4\pm 0.1)$. It is longer than what it is expected from the measured spectral width of $98$ nm, corresponding to $\tau_e\approx 9$ fs. This can be due to a coupling between spatial and frequency modes in the shaper that leads to an effective narrowing of the photons spectrum. The spatial entanglement size is essentially determined by the phase matching in the crystal such that $\sigma_e=(26\pm 1)$ \textmu m \cite{Stefanov2020}. The temporal duration of the classical source is $\tau=15.4$ fs. It is focused into the SFG crystal to a beam waist of $w_0=(5.2\pm 0.2)$ \textmu m or $\sigma=(3.7\pm 0.1)$ \textmu m.

From a quadratic fit of classical SFG signal as a function of the pulsed laser intensity and using Eq. \eqref{eq:ISFGPulsed}, we extract $\beta_c=(6.5 \pm 0.6)\times 10^{-35}$ m$^2$s. Applying Eq. \eqref{eq:betaq_betac} we expect the quantum non-linear coefficient to be $\beta_q^{exp}=(5.5 \pm 0.7)\times 10^{-14}$.

For the quantum source, the down-converted power yields $P_{DC}=(120\pm 4)$ nW, and leads to an up-conversion rate $R_\mathit{UC}=(12.8\pm 0.9)$ Hz (see Fig. \ref{fig:TPA_c2_scan}). Including the setup transmission of $T=62\%$ and the detector efficiency of $\eta=17\%$, we obtain $\beta_q=(4.0 \pm 1.3)\times 10^{-11}$.
The three orders of magnitude discrepancy between the expected and the measured quantum coefficient is mainly due to the fact that we considered $\beta_c$ to be frequency independent. While this would be true for a thin crystal, the phase-matching in the long crystal introduces a strong frequency anti-correlation. This means not all frequency components of the classical pulses contribute equally to the measured signal. By assuming a frequency-independent cross section, we underestimate $\beta_c$ in the relevant frequency region for entangled two-photon absorption. That is why we have to weight $\beta_c$ with an artificially introduced phase-matching factor that quantifies the fraction of frequency components of the incoming pulse $\mathcal{E}(\omega)$ that are used for TPA
\begin{equation}
	p=\frac{\left|\int\mathrm{d}\omega_1\int\mathrm{d}\omega_2\;f_{DC}(\omega_1,\omega_2)\mathcal{E}(\omega_1)\mathcal{E}(\omega_2)\right|}{\left|\int\mathrm{d}\omega_1\int\mathrm{d}\omega_2\;\mathcal{E}(\omega_1)\mathcal{E}(\omega_2)\right|},\quad p\in [0,1],
	\label{eq:TPA_p_artificially}
\end{equation} 
with the phase-matching function $f_{DC}(\omega_i,\omega_s)$. We calculated the acceptance bandwidth of the PPKTP crystal including full transverse dependency. It is not a perfect delta function, but has a finite acceptance bandwidth, that can be approximated by $f_{DC}(\omega_i,\omega_s)\approx\mathrm{exp}\{(\omega_p-\omega_i-\omega_s)^2/(2\Delta\omega_p^2)\}$ with $\Delta\omega_p=0.35$ mrad/fs.
We estimate the error of $\Delta\omega_p$ to be 10\%. In combination with the measured spectrum $S(\omega)$ of the classical source, we approximate the phase-matching factor $p\approx(0.87\pm 0.07)\%$. Finally we can compare the expected quantum cross-section computed from the classical measurement and corrected by the factor $1/p$ with the effectively measured one 

\begin{equation}
\beta_q^{exp}=(6.4 \pm 1.3)\times 10^{-12}, \textrm{ and } \beta_q^{measured}=(40 \pm 13)\times 10^{-12}.
\end{equation}
The two values differs by a factor $6$. This is likely due to the fact that we we didn't took into account the stronger divergence of the classical beams within the crystal. As a consequence we underestimate the classical coefficent.  However the fact that they are on the same order of magnitude supports the model beyond Eq. \eqref{eq:TPA_sigmae}, even when extended to resonant TPA.
 
Because of the formal equivalence between SFG and TPA in molecules, we expect that the entangled photon cross-section of molecules can be, out of any resonance, estimated from Eq. \eqref{eq:betaq_betac}, with $\beta_c$ the classical TPA cross-section and $\beta_q$ the entangled photon ones. Typical values for the TPA cross section $\sigma_c$ are in the range of $10^{-51}$ to $10^{-47}$~cm$^4$s \cite{He2008}, such that the detection of TPA with entangled photon pairs is expected to be bellow the detection limit in conventional materials.

\subsection{Simultaneous Frequency and Time Resolution}
The SLM provides a very flexible way in shaping the two-photon state with some limitations due to its finite frequency resolution. For instance, applying a linear phase as function of frequency, effectively applies a time shift. The smallest possible time shift is limited by the phase shift resolution of the SLM, and is estimated to be 0.007 fs. The largest shift without aliasing effect is constrained by the energy resolution $\Delta\omega=3.3$ mrad/fs, such that $\tau_{max}=2\pi/\Delta\omega=1900$ fs. Indeed, the shaping of classical pulses measured with an autocorrelator (Carpe, APE) shows that for $\tau>2500$ fs, a loss of the signal is observed. 

\begin{figure}[t]
\centering
\includegraphics[width=0.8\columnwidth]{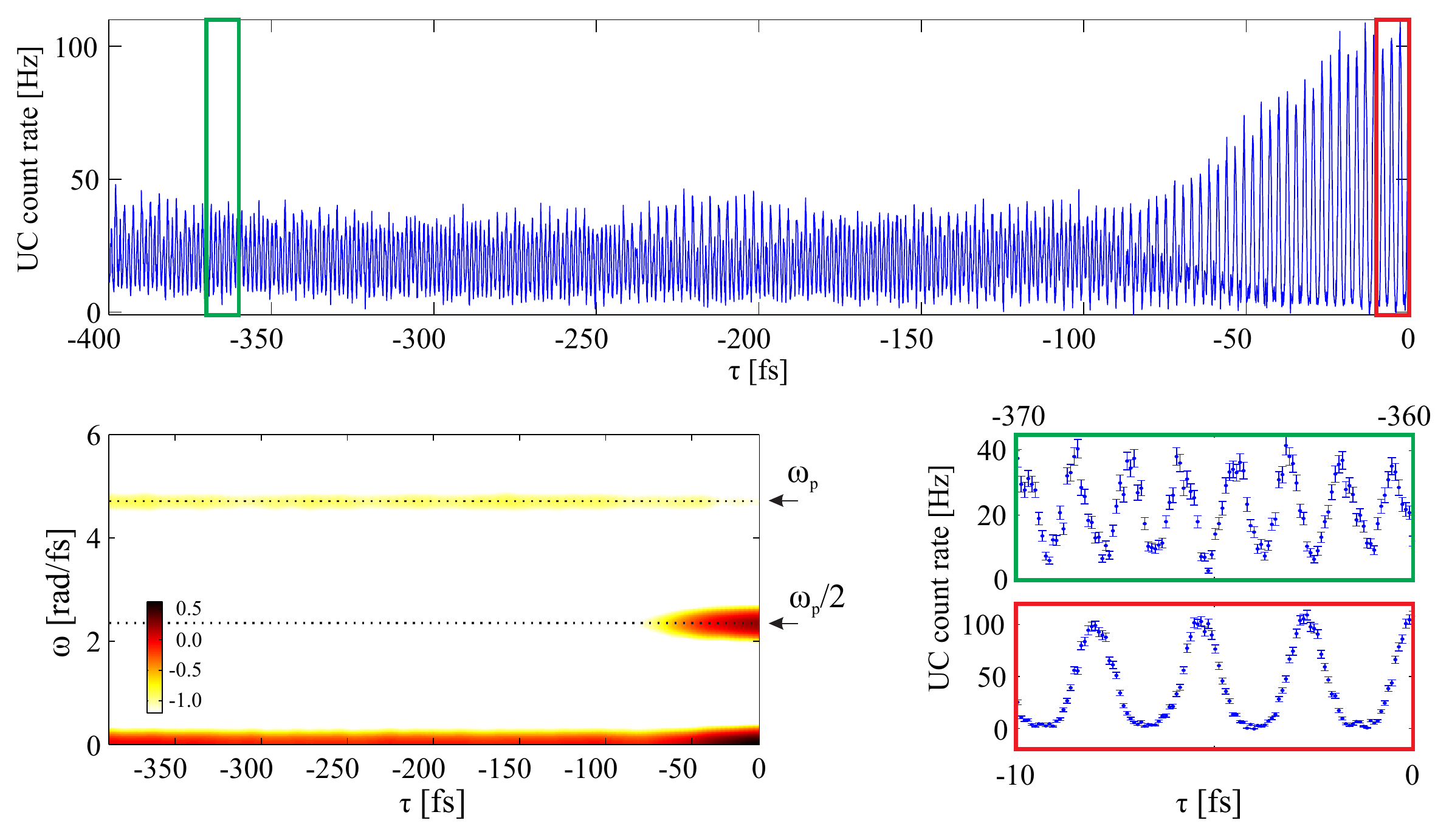}
\caption{Interferometric autocorrelation measurement of the two-photon state. Top plot: dark count ($10.8$~Hz) corrected up-conversion count rate for $9000$ equidistant $\tau$ values in the range from $-400$~fs to $0$~fs. Each count rate is measured during five seconds. The green and red squares indicate the positions of two excerpts that are plotted in the right bottom part of the figure. They illustrate the interferences with different frequencies. The errorbars are calculated assuming Poissonian count statistic. Bottom left: Spectrogram of the measured IAC illustrated in a logarithmic color code. It is a windowed Fourier transform using 256 $\tau$ values (time window of 22.7 fs), and reveals the different oscillation frequencies $\omega_p/2$ and $\omega_p$.}
\label{fig:TPA_IAC}
\end{figure}

We illustrate the capacity to shape entangled photons by exploiting one obvious feature of entangled TPA that is the simultaneous frequency and time resolution. For a two-photon transition to a double-excitation state, the broadband source allows to access a lot of possible pathways to reach the final state (see Fig. \ref{fig:TPAlevelscheme}). At the same time the double-excitation state is spectrally well resolved because of the narrowband pump \cite{Schlawin2012b,Schlawin2013a}. The narrowband and broadband feature of the source is nicely illustrated by means of an interferometric autocorrelation (IAC) measurement \cite{Zah2008}. For that purpose we mimic with the SLM an unbalanced interferometer by means of the transfer function
\begin{equation}
M(\Omega)=\frac{1}{2}\left(1+\mathrm{e}^{-i(\omega_p/2+\Omega)\tau}\right).
\end{equation}
This transfer function acts on both indistinguishable photons. Two types of interferences appears: when signal and idler photon take different paths, they will interfere within their coherence time $\tau_c$ and we expect oscillations with dominating frequency $\omega_p/2$.  If both photons take the same path, the coherence time of the biphoton is equal to the coherence time of the pump photon (380 ns) and interference with frequency $\omega_p$ will be observed even for highly unbalanced interferometer arms. The shaped two-photon states are measured in coincidence by means of the up-conversion crystal. The result is shown in Fig. \ref{fig:TPA_IAC}. For $|\tau|<\tau_c$, we observe oscillations with frequency $\omega_p/2$,  and for $|\tau|>\tau_c$ the oscillations only have frequency $\omega_p$. The fact that signal and idler are broadband leads to a small coherence time $\tau_c$ and the oscillation with $\omega_p/2$ disappear rapidly. 

Even though the dispersion of signal and idler photon has been compensated for this particular measurement, we measure a relatively long coherence time of about 100 fs. This is because the measurement has been performed with a slightly different geometry of the grating compressor which has a worse mode overlap but a similar frequency resolution at the SLM. For time shifts with 3.5 ps, we still observe oscillations that vanish already at 350 fs for a prism compressor setup \cite{Zah2008}. This illustrates that the frequency resolution is a key element for reliable shaping of the two-photon state.

The temporal resolution and accessible time range are thus well suited for performing spectroscopy. For off-resonant entangled two-photon absorption the time needs to be varied within the entanglement time with a resolution of 0.25 fs \cite{Roslyak2009a}. For more complex systems, the lifetimes of intermediate states are on the order of 100~fs \cite{Schlawin2012b}. This time scale can be resolved as long as the entanglement time is shorter.

\section{Conclusion}
\label{sec:CC_conclusion}
In summary, the prism compressor setup and the grating compressor setup allow to create highly energy-time entangled photons by SPDC in a nonlinear crystal. The SLM, established in the field of ultrafast optics, is a convenient tool to manipulate the energy modes of signal and idler photon in a versatile manner. Although both optical arrangements are conceptually equivalent and have to be carefully aligned in order to compensate for dispersion and to guarantee maximal spatial mode overlap, the newly built grating compressor setup is about seven times better in a direct comparison of the frequency resolution. Moreover, its pump frequency can be flexibly tuned in the range of about one nm. A broader tuning range involves realignment of the setup but is principally possible. A major drawback of the new setup compared to the prism compressor setup is the reduced pump power leading to 6.7 times less down-converted photons. A ten times higher pump power or more efficient down-conversion source would lead to similar spectral mode density of 0.2 for both setups. This is still in the single photon limit such that the emitted light can be considered as composed of distinct photon pairs. Hence the only showstopper for the implementation of two-photon spectroscopy with entangled photons is its very low cross-section. Dramatic improvement of the light-matter interaction strength would have thus to be achieved.

\begin{acknowledgments}
This research was supported by the Swiss National Science Foundation through the grant PP00P2\textunderscore159259. 
\end{acknowledgments}

\appendix
\section{Shapper alignment}
\label{app:alignement}
\begin{figure}[htbp]
\centering
\includegraphics[width=0.5\linewidth]{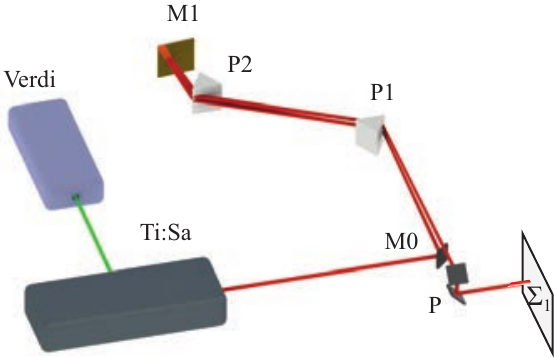}
\caption{Classical broadband source. A Ti:Sa oscillator generates femtosecond pulses. The pulses are sent via mirror M0 through a prism compressor (prisms P1 \& P2, mirror M1) to obtain nearly transform-limited pulses. After a periscope (P) that rotates the polarization by $90^\circ$, the pulses are coupled into the manipulation stage at $\Sigma_1$.}
\label{fig:GC_generation}
\end{figure}

The alignment of the shapper is realized with a classical broadband source depicted in Fig. \ref{fig:GC_generation}. A 5 W Verdi laser pumps an Ti:Sa oscillator (KM Labs, Model MTS). The oscillator emits $46$ nm broad pulses centered around $\lambda_c=800$ nm at a repetition rate of $\nu=90$ MHz.
The generated pulses are not transform-limited. The output mirror in the oscillator and the free propagation of $4.4$ m through air introduce dispersion to the pulses before they are coupled into the manipulation stage at $\Sigma_1$. A prism compressor, composed of two N-SF11 equilateral prisms (P1 and P2), compensates for this dispersion. A periscope (P) manipulates the beam's polarization from p-polarization to s-polarization in order to guarantee the phase-matching condition at the detection crystal. The beam travels $4.4$ m distance from the oscillator output to the manipulation stage at $\Sigma_1$, and can be considered as collimated.

The coincidence detection of signal and idler by up-conversion and the SFG detection of classical laser pulses are very sensitive to dispersion. While propagating through the setup, each photon picks up the same additional phase $\phi(\Omega)$. In a perfect $4f_2$ configuration, the grating compressor introduces zero dispersion. Additional phases are caused mainly by the down- and up-conversion crystal, and the lenses L1 and L2. These phases precludes coincidence detection of signal and idler at $\Sigma_4$. When shifting the grating G1 (G2) from $\Sigma_1$ ($\Sigma_3$) to $\Sigma_0$ ($\Sigma_4$) by the distance $g$, the setup compensates ($g>0$) or introduces ($g<0$) dispersion \cite{Konrad2014}. Therefore, it is possible to compensate $\phi(\Omega)$, by choosing $g$ adequately.

The dispersion sensitivity of the detection depending on $g$ is probed with the classical source. While grating G1 is at fixed position, G2 is shifted. The incoming pulses $\mathcal{E}_0^+(\Omega)$ pick up the additional phase of the setup, and are shaped by the SLM transfer function
\begin{equation}
M(\Omega)=\mathrm{e}^{i\frac{c_2'}{2}\Omega^2}
\label{eq:M_c2}
\end{equation}
such that they read
\begin{equation}
\mathcal{E}^+(\Omega)=\mathcal{E}_0^+(\Omega)M(\Omega)\mathrm{e}^{i\phi(\Omega)}.
\end{equation}
The phase can be expanded in a Taylor series
\begin{equation}
\phi(\Omega)=\sum\limits_{k=0}^\infty \frac{c_k}{k!}\Omega^k= c_0+c_1\Omega+\frac{c_2}{2}\Omega^2+\mathcal{O}(\Omega^3),
\label{eq:CC_phi_Taylor}
\end{equation}
with the Taylor coefficients
\begin{equation}
c_k=\left.\frac{\partial^k\phi}{\partial\omega^k}\right|_{\omega=\omega_c}.
\end{equation}
A photodiode after the up-conversion crystal detects the signal 
\begin{eqnarray}
S_\textit{PD}&=&\int\mathrm{d}\Omega_3\;|S(\Omega_3)|^2\nonumber\\
&\propto&\int\mathrm{d}\Omega_3\left|\int\mathrm{d}\Omega\;\mathcal{E}^+(\Omega)\mathcal{E}^+(\Omega_3-\Omega)f_\textit{UC}(\Omega_3-\Omega,\Omega)\right|^2\nonumber\\
&\propto&\int\mathrm{d}\Omega_3\left|\int\mathrm{d}\Omega\;\mathcal{E}_0^+(\Omega)\mathcal{E}_0^+(\Omega_3-\Omega)f_\textit{UC}(\Omega_3-\Omega,\Omega)\mathrm{e}^{i(c_2'+c_2)\Omega^2}\right|^2.
\label{eq:CC_S_photodiode}
\end{eqnarray}
For transform-limited input pulses, the maximal signal is measured when the SLM compensates the setup's group velocity dispersion (GVD) coefficient, i.e. $c_2'=-c_2$.

\begin{figure}[t]
\centering
\includegraphics[width=0.6\linewidth]{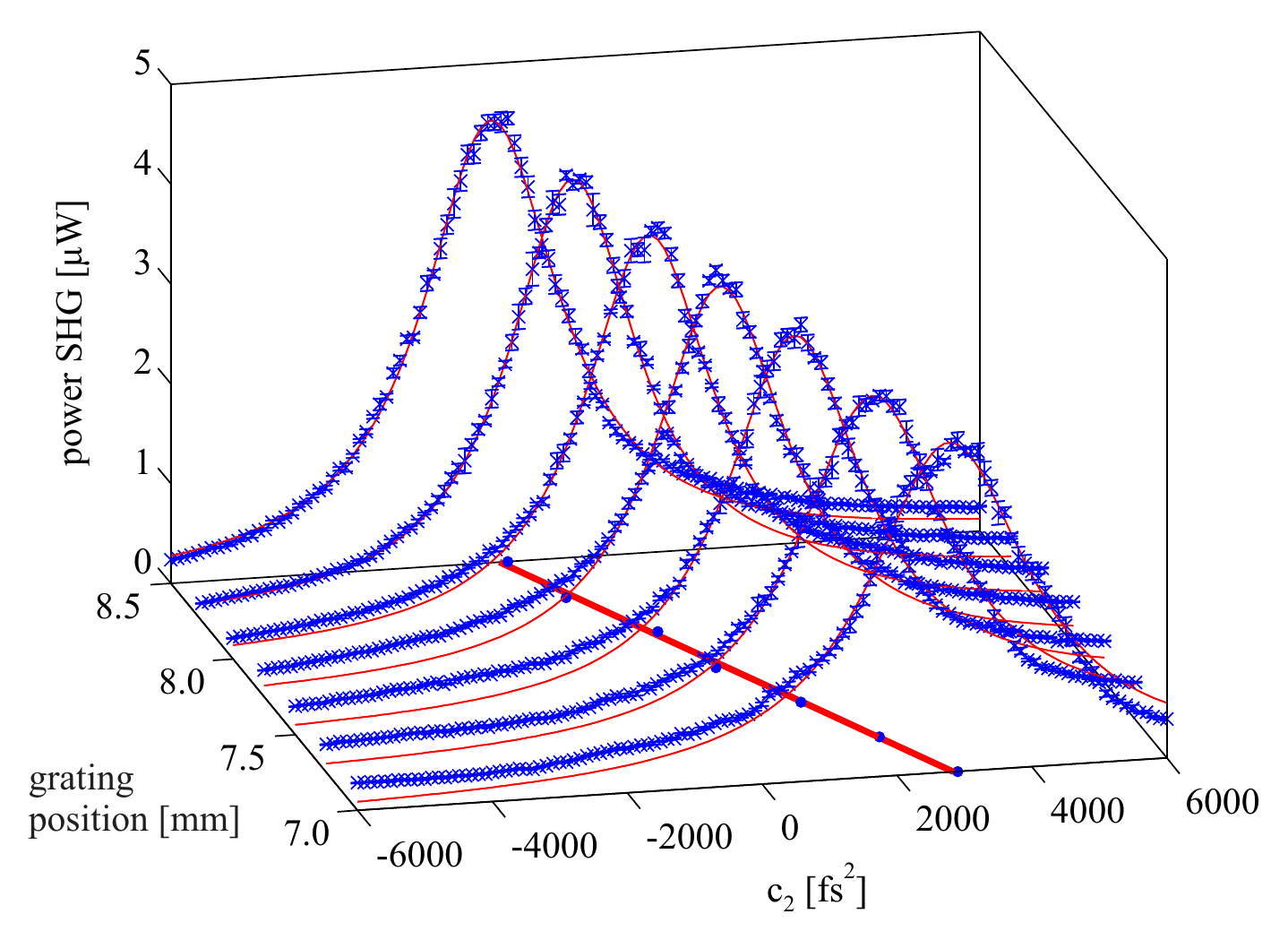}
\caption{GVD scans for different relative grating positions (blue crosses). Each scan is fitted by a Lorentzian function (red curve). For each fit the maximum position is projected onto the plane at zero power (blue circles). The circles follow a linear slope (red line) of $(-2643\pm92)$ fs$^2$/mm which coincides with the ZEMAX simulation of the setup.}
\label{fig:GC_dispersion_scan}
\end{figure}

A measurement for seven different grating positions is depicted in Fig. \ref{fig:GC_dispersion_scan}. For each value of $c_2'$ and each grating position, the photodiode measures the SFG signal $30$ times with a time interval of $10$ ms. A Lorentzian function is fitted using 41 measurement points around the maximum. The maximum positions are then linearly fitted with a slope of $(-2643\pm92)$ fs$^2$/mm. A simulation with the ray tracing software ZEMAX (Version January, 2003) gives $(-2610\pm5)$ fs$^2$/mm. It calculates the frequency-dependent optical path $L_{opt}(\Omega)$ of the beam through the setup up to the center of the UC crystal. With this information, the phase $\phi(\Omega)=2\pi L_{opt}(\Omega)/\lambda_c$ can be used for the Taylor series in Eq. \eqref{eq:CC_phi_Taylor}. For the simulation, higher order terms can be included too, but those change the slope only minimally within its error.

\begin{figure}[t]
\centering
\includegraphics[width=0.9\linewidth]{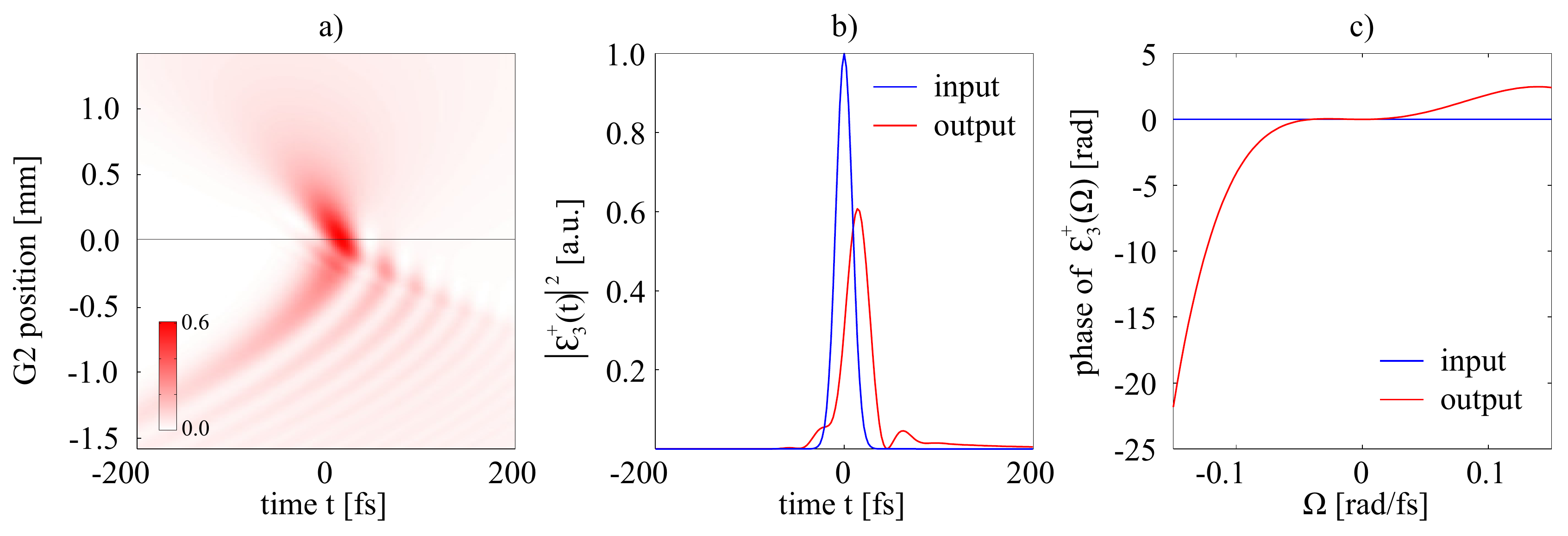}
\caption{a) Temporal profile $\left|\mathcal{E}^+(t)\right|^2$ of the outgoing pulse for different relative positions of grating G2. Grating G1 is at fixed position such that the setup is symmetric for the relative position zero. The profile is normalized to the maximum of the incoming pulse. b) Comparison of the temporal profile $\left|\mathcal{E}^+(t)\right|^2$ of the incoming pulse (blue) and the outgoing pulse with minimal width at relative grating position zero (red). c) While the phase of the transform-limited pulse (blue) is zero per definition, the residual phase of the outgoing pulse cannot be fully compensated with accurate grating positions (red).}
\label{fig:GC_dispersion_theory}
\end{figure}

The simulation shows not only GVD but also higher order dispersion terms. There are no grating positions such that the setup fully compensates those terms. Figure \ref{fig:GC_dispersion_theory} a) illustrates this. For different positions of G2, the temporal shape of an incoming transform-limited $12.3$ fs pulse after propagation through the setup is plotted. At the relative grating position zero, the pulse has minimal width but is not transform-limited. Fig.~\ref{fig:GC_dispersion_theory} b) shows the slice plane at this position. The residual phase is plotted in Fig.~\ref{fig:GC_dispersion_theory} c). It can easily be compensated with the SLM. For other positions of grating G1, the plots look similar, but the absolute position of G2 changes.

%
\end{document}